\documentstyle[prl,aps,epsf,amsmath,amssymb]{revtex}

\newcommand{\bea}{\begin{eqnarray}}
\newcommand{\eea}{\end{eqnarray}}
\newcommand{\beq}{\begin{equation}}
\newcommand{\eeq}{\end{equation}}
\newcommand{\del}{\partial}
\newcommand{\lishi}{\langle\!\langle}
\newcommand{\rishi}{\rangle\!\rangle}

\begin{document}

%%%%%%%%%%%%%%%%%%%%%%%%%%%%%%%%%%%%%%%%%%%%%%%%%%%%%%%%%%%%%%%%%%

%%% \draft command makes pacs numbers print %%%
\draft

%%% Comment out following 2 lines (& 1 more below) if 1-column %%%
\twocolumn[\hsize\textwidth\columnwidth\hsize\csname
@twocolumnfalse\endcsname

\preprint{OUTP-2002-39P, HIP-2002-45/TH, hep-th/0210032}

\title{Free-field realisation of boundary states and boundary
correlation functions of minimal models}
% Force line breaks with \\

\author{Shinsuke Kawai}
\address{%
Theoretical Physics, Department of Physics, University of Oxford, 
1 Keble Road, Oxford OX1 3NP, UK, and
}%
\address{%
Helsinki Institute of Physics, 
P.O. Box 64, %(Gustaf H\"allstr\"omin katu 2)
FIN-00014 University of Helsinki, Finland
}%

\date{\today}
\maketitle

%%%%%%%%%%%%%%%%%%%%%%%%%%%%%%%%%%%%%%%%%%%%%%%%%%%%%%%%%%%%%%%%%%
%%%%% Abstract %%%%%
\begin{abstract}
We propose a general formalism to compute exact correlation functions
for Cardy's boundary states.
Using the free-field construction of boundary states and applying the Coulomb-gas
technique, it is shown that charge-neutrality conditions pick up particular linear 
combinations of conformal blocks.
As an example we study the critical Ising model with free and fixed boundary
conditions, and demonstrate that conventional results are reproduced.
This formalism thus directly associates algebraically constructed
boundary states with correlation functions which are in principle 
observable or numerically calculable.
\end{abstract}
%%%%%%%%%%%%%%%%%%%%%%%%%%%%%%%%%%%%%%%%%%%%%%%%%%%%%%%%%%%%%%%%%%%

% PACS, the Physics and Astronomy Classification Scheme.
%\keywords{boundary conformal field theory, Coulomb-gas formalism}
%Use showkeys class option if keyword display desired

%%% Comment out if one column %%%
\vskip2pc]

%%%%%%%%%%%%%%%%%%%%%%%%%%%%%%%%%%%%%%%%%%%%%%%%%%%%%%%%%%%%%%%%%%%
%%%%% Body of the Paper %%%%%

\section{Introduction}
Modular invariance of partition functions plays extremely important roles in
two-dimensional conformal field theory (CFT). 
The ADE classification of modular invariants by Cappelli, Itzykson and Zuber 
\cite{ADEmin,ADEsu2} (see also \cite{ADEgalois,ADEsu3,ADEN=2}) is obtained by 
considering CFTs on the torus. 
The classified modular invariants correspond to particular sets of operators, 
which are supposed to model critical systems in certain universality classes.
Similar consideration also applies to CFTs defined on a manifold with 
boundary.
For a CFT on the cylinder, the constraints from modular invariance lead to a classification of boundary states. 
This method was invented by Cardy\cite{cardymodular} in the eighties 
and developed by many in the 
nineties\cite{cardylewellen,lewellen,pradisi,runkel,bppz,fsch}. 
Such a classification of boundary states with consistent modular properties 
has recently attracted much attention along with the development of D-brane / 
string theory and various application of boundary CFT to statistical physics. 

Boundary states with consistent modular properties, or {\em consistent}
boundary states, are normally believed to represent boundary conditions which
may be physically imposed on D-branes or borders of statistical systems. 
In order to understand the behaviour of observables in the presence of such a boundary,
we need to find correlation functions for consistent boundary states which are defined through Cardy's classification. 
In principle this can be accomplished by operator product expansions (OPEs), that is,
by finding boundary operators for a given boundary state, 
solving the constraints satisfied by coupling constants,
and then obtaining the correlation functions by OPEs using the boundary operators. 
The correlation functions obtained in this way are perturbative, that is, in the form of series expansion.
For practical use, we often need to know exact correlation functions and in such 
a case we normally solve differential equations to find conformal blocks and fix their coefficients by physical considerations\cite{cardycorrel}. 
In this differential equation method, however, the relation to Cardy's 
classification of boundary states is not quite evident.

In this paper, we present an alternative formalism for finding boundary
correlation functions\footnote{We use this term for correlation functions of {\em bulk} operators
in the presence of boundary.} directly using boundary states obtained by Cardy's 
method. 
It is well known in string theory that correlation functions are 
simply given by inserting vertex operators within amplitudes (with or 
without boundaries). 
This picture is generalised to non-bosonic ($c\neq 1$) CFTs using the 
Coulomb-gas formalism on arbitrary Riemann surfaces (without boundaries)
\cite{felder,FelSil,DiPFHLS,flms}. 
The purpose of this paper is to present a method to calculate boundary 
correlation functions based on the Coulomb-gas picture, by using the
free-field representation of boundary states developed in \cite{coulomb}.
Driven by a similar motivation, Coulomb-gas system on the half plane is discussed in
\cite{schulze}, where the Ising model conformal blocks are reproduced using the contour integration technique of \cite{mathur} and insertion of boundary operators.
The key ingredient of the formalism proposed in the present paper is
boundary states having boundary charges, which account {\em both} for the contour integral expression of conformal blocks {\em and} for their coefficients.
The role of such boundary states has not been fully investigated so far in this context,
and our formalism allows a systematic study on the relation between correlation functions and algebraically defined boundary states. 
In the following we shall mainly consider Coulomb-gas systems on the unit disk, where Felder's 
charged bosonic Fock space (CBFS) construction\cite{felder} is readily used.
Although our method itself is quite general, we shall focus on presenting the 
ideas in simplest cases and show that it reproduces known results obtained
by the conventional approach.

We organise the rest of this paper as follows. 
In the next section we fix our notations and 
review the free-field construction of boundary states\cite{coulomb}. 
We describe in Sec. III our method to compute correlation functions on the disk
and on the half plane. 
In Sec. IV we illustrate the method using the Ising model and show that 
it reproduces the results of \cite{cardycorrel}. 
Finally in Sec. V we summarise and conclude.

%%%%%%%%%%%%%%%%%%%%%%%%%%%%%%%%%%%%%%%%%%%%%%%%%%%%%%%%%%%%%%%%%%%

\section{Boundary states in Coulomb-gas formalism}

Let us start, for the sake of self-containedness, by summarising the
Coulomb-gas construction of boundary states \cite{coulomb}.
The idea is to define coherent states in the charged bosonic Fock spaces 
(CBFSs) and find conditions for their diffeomorphism invariance and modular
consistency. 

\subsection{Coulomb-gas and charged bosonic Fock space}

In the Coulomb-gas formalism\cite{dotfat}, the Virasoro minimal models are 
realised by the action,
\beq
{\cal S}=\frac{1}{8\pi}\int d^2x\sqrt g(\del_\mu\Phi\del^\mu\Phi
+2\sqrt 2\alpha_0i\Phi R),
\label{eqn:action}
\eeq
where the scalar field $\Phi(x)$ is assumed to decouple into two chiral 
sectors,
$\Phi(z,\bar z)=\varphi(z)+\bar\varphi(\bar z)$. 
The energy-momentum tensor is obtained by the variation of the action 
(\ref{eqn:action}) as
\beq
T(z)=-2\pi T_{zz}
=-\frac 12 :\del\varphi\del\varphi:+i\sqrt 2\alpha_0\del^2\varphi.
\eeq
The central charge of this system is 
\beq
c=1-24\alpha_0^2.
\label{eqn:central}
\eeq
The vertex operators,
\beq
V_\alpha(z)=:e^{i\sqrt 2\alpha\varphi(z)}:,
\eeq
are chiral fields of conformal dimensions
$
h_\alpha=\alpha^2-2\alpha_0\alpha.
$
In particular, the primary fields $\phi_{r,s}$ ($0<r<p'$, $0<s<p$) of a minimal 
model are realised by the vertex operators $V_{\alpha_{r,s}}(z)$ with
\beq
\alpha_{r,s}=\frac 12(1-r)\alpha_++\frac 12(1-s)\alpha_-, 
\label{eqn:minimalcharge}
\eeq
where
$\alpha_+=\sqrt{p/p'}$, $\alpha_-=-\sqrt{p'/p}$, and
$p$ and $p'$ (we assume $p>p'$) are the two coprime integers characterising 
the minimal model. 
% $\alpha_{\pm}=\alpha_0\pm\sqrt{\alpha_0^2+1}$.
The conformal dimensions of the operators are
\beq
h_{r,s}=\frac 14(r\alpha_++s\alpha_-)^2-\alpha_0^2,
\eeq
which agree with the Kac formula.

The holomorphic chiral boson field is expanded in modes as
\beq
\varphi(z)=\varphi_0-ia_0\ln z+i\sum_{n\neq 0}\frac{a_n}{n}z^{-n},
\label{eqn:phimode}
\eeq
and likewise for the antiholomorphic counterpart,
\beq
\bar\varphi(\bar z)=\bar\varphi_0-i\bar a_0\ln\bar z
+i\sum_{n\neq 0}\frac{\bar a_n}{n}\bar z^{-n}.
\label{eqn:barphimode}
\eeq
As we try to consider boundary CFT in the Coulomb-gas picture,
a subtlety arises in the treatment of zero-mode, since the zero-mode of
$\Phi(z,\bar z)$ does not naturally decouple into the holomorphic and 
antiholomorphic sectors.
In our formalism, we shall simply split it into two identical and independent 
copies. 
In this sense, the boundary theory we shall consider is not exactly a 
non-chiral theory on a manifold with boundary but rather a chiral theory on 
its Schottky double\footnote{The Schottky double is a Riemann surface obtained by 
doubling the manifold except for boundaries. See, e.g.,  \cite{sfw}.}.
We thus have two copies of Heisenberg operators, satisfying the algebra
\bea
&&[a_m,a_n]=m\delta_{m+n,0},\;\;\;[\varphi_0,a_0]=i,\nonumber\\
&&[\bar a_m,\bar a_n]=m\delta_{m+n,0},\;\;\;[\bar\varphi_0,\bar a_0]=i,
\label{eqn:heisenberg}
\eea
with no interaction,
\beq
[a_m,\bar a_n]=0,\;\;\;[\varphi_0,\bar a_0]=[\bar\varphi_0,a_0]=[\varphi_0,\bar\varphi_0]=0.
\eeq
In terms of the Heisenberg operators, the Virasoro operators are written as
\bea
&&L_{n\neq 0}=\frac 12\sum_{k\in{\mathbb Z}}a_{n-k}a_k-\sqrt 2\alpha_0(n+1)a_n,
\label{eqn:virasoron}\\
&&L_0=\sum_{k\geq1}a_{-k}a_k+\frac 12 a_0^2-\sqrt 2\alpha_0a_0,
\label{eqn:virasoro0}
\eea
and likewise for the antiholomorphic counterparts.

The Hilbert space of CFT is realised in CBFSs with BRST projection
\cite{felder}.
For a {\em chiral} theory the CBFS $F_{\alpha;\alpha_0}$ is defined as a space 
obtained by operating with the creation operators $a_{n<0}$ on the chiral 
highest weight state $\vert\alpha;\alpha_0\rangle$ 
(see \cite{felder,coulomb,dFMS}).
Since the boundary intertwines the two chiral sectors, we need to construct
non-chiral Fock spaces ${\cal F}_{\alpha,\bar\alpha;\alpha_0}$ in order to 
describe boundary states. 
We denote the M\"obius invariant non-chiral vacua with background charge 
$\alpha_0$ as\footnote{Here we give the same 
background charge $\alpha_0$ to both sectors. 
Even if we relax this condition and start by allocating different 
background charges $\alpha_0$ and $\bar\alpha_0$ to holomorphic and 
antiholomorphic sectors, respectively, the condition (\ref{eqn:ishibashicond}) 
restricts either $\alpha_0=\pm\bar\alpha_0$. For $\alpha_0=-\bar\alpha_0$ we
have $\Omega=-1$ and $\alpha-\bar\alpha-2\alpha_0=0$ instead of
(\ref{eqn:omega}) and (\ref{eqn:alpha}). 
This merely flips the sign of all antiholomorphic charges and thus does not 
give any new results.}
$\langle 0,0;\alpha_0\vert$ and $\vert 0,0;\alpha_0\rangle$,
and define highest weight vectors $\langle\alpha,\bar\alpha;\alpha_0\vert$ and 
$\vert\alpha,\bar\alpha;\alpha_0\rangle$ as
\bea
&&\langle\alpha,\bar\alpha;\alpha_0\vert
=\langle 0,0;\alpha_0\vert 
e^{-i\sqrt 2\alpha\varphi_0}e^{-i\sqrt 2\bar\alpha\bar\varphi_0},\\
&&\vert\alpha,\bar\alpha;\alpha_0\rangle
=e^{i\sqrt 2\alpha\varphi_0}e^{i\sqrt 2\bar\alpha\bar\varphi_0}
\vert 0,0;\alpha_0\rangle.
\eea
The state $\langle\alpha,\bar\alpha;\alpha_0\vert$ has holomorphic and
antiholomorphic charges $-\alpha$ and $-\bar\alpha$, respectively.
Likewise, $\vert\alpha,\bar\alpha;\alpha_0\rangle$ has holomorphic and 
antiholomorphic charges $\alpha$ and $\bar\alpha$. 
Using the Heisenberg algebra (\ref{eqn:heisenberg}) it is easy to verify that
these states satisfy
\bea
&&\langle\alpha,\bar\alpha;\alpha_0\vert a_0
=\langle\alpha,\bar\alpha;\alpha_0\vert\sqrt 2\alpha,
\label{eqn:zeromodeop1}\\
&&\langle\alpha,\bar\alpha;\alpha_0\vert\bar a_0
=\langle\alpha,\bar\alpha;\alpha_0\vert\sqrt 2\bar\alpha,
\label{eqn:zeromodeop2}\\
&&a_0\vert\alpha,\bar\alpha;\alpha_0\rangle
=\sqrt 2\alpha\vert\alpha,\bar\alpha;\alpha_0\rangle,
\label{eqn:zeromodeop3}\\
&&\bar a_0\vert\alpha,\bar\alpha;\alpha_0\rangle
=\sqrt 2\bar\alpha\vert\alpha,\bar\alpha;\alpha_0\rangle.
\label{eqn:zeromodeop4}
\eea
These highest weight vectors are eigen states of the Virasoro zero-modes:
\bea
L_0\vert\alpha,\bar\alpha;\alpha_0\rangle
=(\alpha^2-2\alpha\alpha_0)\vert\alpha,\bar\alpha;\alpha_0\rangle,\\
\bar L_0\vert\alpha,\bar\alpha;\alpha_0\rangle
=(\bar\alpha^2-2\bar\alpha\alpha_0)
\vert\alpha,\bar\alpha;\alpha_0\rangle.
\eea
The states $\vert\alpha,\bar\alpha;\alpha_0\rangle$ are annihilated by
$a_{n>0}$, $\bar a_{n>0}$, $L_{n>0}$, $\bar L_{n>0}$, and 
$\langle\alpha,\bar\alpha;\alpha_0\vert$ are annihilated by 
$a_{n<0}$, $\bar a_{n<0}$, $L_{n<0}$, $\bar L_{n<0}$.
The non-chiral CBFS ${\cal F}_{\alpha,\bar\alpha;\alpha_0}$ is built on the
highest weight vector $\vert\alpha,\bar\alpha;\alpha_0\rangle$ by operating
with $a_{n<0}$ and $\bar a_{n<0}$. 
The dual space ${\cal F}_{\alpha,\bar\alpha;\alpha_0}^*$ is defined similarly, 
by acting with $a_{n>0}$ and $\bar a_{n>0}$ on 
$\langle\alpha,\bar\alpha;\alpha_0\vert$.
The non-chiral CBFSs thus defined are essentially the direct products of
chiral CBFSs, ${\cal F}_{\alpha,\bar\alpha;\alpha_0}=F_{\alpha;\alpha_0}\otimes
\bar F_{\bar\alpha;\alpha_0}$.

The inner products of highest weight vectors are subject to charge
neutrality, i.e. they are non-vanishing only if the net charges in the two
sectors are both zero. 
The normalisation of the highest weight vectors must be in accordance with 
this condition and thus we have
\beq
\langle\alpha,\bar\alpha;\alpha_0\vert\beta,\bar\beta;\alpha_0\rangle
=\kappa\delta_{\alpha,\beta}\delta_{\bar\alpha,\bar\beta}.
\label{eqn:hwvnorm}
\eeq
In particular, the vacua are normalised as
\beq
\langle0,0;\alpha_0\vert 0,0;\alpha_0\rangle=\kappa.
\label{eqn:vacnorm}
\eeq
In unitary theories the constant $\kappa$ is usually positive and we normalise
it to unity.
As the Coulomb-gas system may well include non-unitary theories, $\kappa$ can 
be negative. In that case we choose $\kappa=-1$. 
Thus we have
\bea
&&{}_U\langle\alpha,\bar\alpha;\alpha_0\vert\beta,\bar\beta;\alpha_0\rangle_U
=\delta_{\alpha,\beta}\delta_{\bar\alpha,\bar\beta}
\label{eqn:hwvnormU},\\
&&{}_N\langle\alpha,\bar\alpha;\alpha_0\vert\beta,\bar\beta;\alpha_0\rangle_N
=-\delta_{\alpha,\beta}\delta_{\bar\alpha,\bar\beta}
\label{eqn:hwvnormN}.
\eea
The subscripts $U$ and $N$ stand for unitary and non-unitary sectors, 
respectively.
Those two sectors have no intersection.

\subsection{Diffeomorphism invariant boundary states}

Boundary states appearing in CFT are diffeomorphism invariant in the following
sense\cite{cardymodular,ishibashi}.
Let us consider a CFT on the upper half plane ${\rm Im} \zeta\geq 0$,
where $\zeta$ is a complex coordinate, $\zeta=x+iy$, $x$, $y\in{\mathbb R}$.
The boundary is $y=0$, or $\zeta=\zeta^*$.
Since the antiholomorphic coordinate dependence $\bar\zeta$ on the upper half 
plane may be mapped into the holomorphic dependence $\zeta^*$ on the lower 
half plane \cite{cardycorrel}, we often identify $\bar\zeta$ with $\zeta^*$. 
Now, once the boundary is fixed, the conformal symmetry of the theory 
should be restricted so that the boundary is kept fixed. 
In other words, as the conformal transformation is generated by the 
energy-momentum tensor, the energy-momentum flow across the boundary must 
vanish,
\beq
\left[T(\zeta)-\bar T(\bar\zeta)\right]_{\zeta=\bar\zeta}=0.
\label{eqn:diffinvT}
\eeq
This may be translated into a condition on boundary states by mapping a 
semiannular domain on the upper half $\zeta$-plane into a full-annulus on the 
$z$-plane by $z=\exp(-2\pi i\xi/L)$, $\xi=(T/\pi)\ln\zeta$. 
The boundary of the $\zeta$-plane is mapped to the two concentric circles
$\vert z\vert=1$, $\exp(2\pi T/L)$ bordering the annulus on the $z$-plane.
Since the $z$-plane allows radial quantisation, (\ref{eqn:diffinvT}) is written
as the condition on the boundary states $\vert B\rangle$ (on $\vert z\vert=1$),
\beq
(L_k-\bar L_{-k})\vert B\rangle=0.
\label{eqn:ishibashicond}
\eeq
This condition, often called the Ishibashi condition, must then be satisfied by
any boundary state in CFT.

We may follow the standard construction of boundary states in open string
theory\cite{clny,polcai,ishiono} and find boundary states on CBFS by starting 
from the coherent state ansatz,
\bea
&&{}_{\kappa}\langle B_{\alpha,\bar\alpha;\alpha_0;\Omega}\vert
={}_{\kappa}\langle\alpha,\bar\alpha;\alpha_0\vert\prod_{k>0}
\exp\left(-\frac{1}{k\Omega}a_k\bar a_k\right),\nonumber\\
\label{eqn:bbra}\\
&&\vert B_{\alpha,\bar\alpha;\alpha_0;\Omega}\rangle_{\kappa}
=\prod_{k>0}\exp\left(-\frac{\Omega}{k}a_{-k}\bar a_{-k}\right)
\vert\alpha,\bar\alpha;\alpha_0\rangle_{\kappa}.\nonumber\\
\label{eqn:bket}
\eea
The subscript $\kappa$ is either $U$ or $N$, specifying the normalisation of 
the vacuum.
As we have expressions of the Virasoro operators
(\ref{eqn:virasoron}), (\ref{eqn:virasoro0}) written in terms of the
Heisenberg operators, one can see how the Virasoro modes $L_{n}$ and 
$\bar L_{-n}$ operate on the coherent states 
$\vert B_{\alpha,\bar\alpha;\alpha_0;\Omega}\rangle_\kappa$
by an explicit computation. 
It is shown \cite{coulomb} that the condition (\ref{eqn:ishibashicond}) is 
satisfied if
\beq
\Omega=1,
\label{eqn:omega}
\eeq
and 
\beq
\alpha+\bar\alpha-2\alpha_0=0.
\label{eqn:alpha}
\eeq
Similarly, we see that 
${}_\kappa\langle B_{\alpha,\bar\alpha;\alpha_0;\Omega}\vert
(L_n-\bar L_{-n})=0$
as long as (\ref{eqn:omega}) and (\ref{eqn:alpha}).
In the following we shall only consider such manifestly diffeomorphism 
invariant boundary states satisfying (\ref{eqn:omega}) and (\ref{eqn:alpha}),
and for simplicity we denote,
\bea
&&{}_{\kappa}\langle B(\alpha)\vert
={}_{\kappa}\langle B_{\alpha,2\alpha_0-\alpha;\alpha_0;\Omega=1}\vert,
\label{eqn:cohbsbra}\\
&&\vert B(\alpha)\rangle_{\kappa}
=\vert B_{\alpha,2\alpha_0-\alpha;\alpha_0;\Omega=1}\rangle_{\kappa}.
\label{eqn:cohbsket}
\eea  

We note that the sum of the (holomorphic $+$ antiholomorphic) boundary 
charges agrees with the topological background charge on the Schottky double.
Due to the condition (\ref{eqn:alpha}), an inner boundary (on the $z$-plane)
contributes $2\alpha_0$ to the sum of the charges, and an outer boundary 
contributes $-2\alpha_0$. 
When we consider an annulus whose Schottky double is a torus, the sum of the 
boundary charges is zero ($2\alpha_0-2\alpha_0=0$), which coincides with
the background charge of the torus expected from the Gauss-Bonnet theorem
(the Euler number of a torus vanishes). 
For a disk, there is only an outer boundary which gives a charge $-2\alpha_0$.
This agrees with the background charge of a sphere, which is the Schottky
double of the disk.
Thus, the geometry of bulk can be assumed to be flat everywhere, since the 
curvature of the manifold is concentrated on boundary.

\subsection{Ishibashi states in free-field representation}

As the basis of boundary states in CFT is normally spanned by Ishibashi states,
we need to construct Ishibashi states in terms of our Fock space representation in order to
translate existing results of boundary CFT into Coulomb-gas language.
At least for the diagonal minimal models, it is shown \cite{coulomb} that  
the Ishibashi states are expressible as linear combinations of 
$\vert B(\alpha)\rangle_{\kappa}$,
as far as partition functions on the cylinder are concerned.

Ishibashi states are defined for chiral representations of CFT and 
diagonalise the cylinder amplitudes (overlaps) to give characters,
\beq
\lishi i\vert(\tilde q^{1/2})^{L_0+\bar L_0-c/12}\vert j\rishi
=\delta_{ij}\chi_j(\tilde q).
\label{eqn:ishiamp}
\eeq
Here, we are considering a cylinder of length $T$ and circumference $L$, or 
equivalently, an annulus on the $z$-plane with 
$1\leq\vert z\vert\leq\exp(2\pi T/L)$. 
As the Hamiltonian is written as $H=(2\pi/L)(L_0+\bar L_0-c/12)$,
the left hand side of (\ref{eqn:ishiamp}) is
$\lishi i\vert e^{-TH}\vert j\rishi$ with $\tilde q=e^{-4\pi T/L}$.
The characters of the minimal models are given by Rocha-Caridi \cite{rocha},
\bea
\chi_{(r,s)}(q)&=&{\rm Tr}_{(r,s)}q^{L_0-c/24}\nonumber\\
&=&\frac{\Theta_{pr-p's,pp'}(\tau)-\Theta_{pr+p's,pp'}(\tau)}{\eta(\tau)},
\label{eqn:rocha}
\eea
where $\Theta_{\lambda,\mu}(\tau)$ and $\eta(\tau)$ are the Jacobi theta
function and the Dedekind eta function, defined as
$\Theta_{\lambda,\mu}(\tau)\equiv\sum_{k\in{\mathbb Z}}
q^{(2\mu k+\lambda)^2/4\mu}$ and
$\eta(\tau)\equiv q^{1/24}\prod_{n\geq1}(1-q^n)$, with 
$q=e^{2\pi i\tau}$.
Thus, the cylinder amplitudes (\ref{eqn:ishiamp}) are power series in 
$\tilde q$, divided by $\eta(\tilde\tau)$.

As our boundary states $\vert B(\alpha)\rangle_{\kappa}$ are defined in a
Fock space representation, we may explicitly compute the cylinder amplitudes
between such states. 
They are\cite{coulomb},
\bea
&&{}_{\kappa}\langle B(\alpha)\vert e^{-TH}\vert B(\beta)\rangle_{\kappa}
\nonumber\\
&=&{}_{\kappa}\langle B(\alpha)\vert(\tilde q^{1/2})^{L_0+\bar L_0-c/12}
\vert B(\beta)\rangle_{\kappa}\nonumber\\
&=&\frac{\tilde q^{(\alpha-\alpha_0)^2}}{\eta(\tilde\tau)}\kappa
\delta_{\alpha,\beta}.
\label{eqn:cbfsamp}
\eea
The amplitudes between unitary and non-unitary sectors 
(e.g. ${}_{U}\langle B(\alpha)\vert e^{-TH}\vert B(\beta)\rangle_{N}$)
vanish because these sectors do not intersect.
Note that $q^{(\alpha-\alpha_0)^2}/\eta(\tau)$ is the character
$\chi_{\alpha;\alpha_0}(q)$ of the chiral CBFS $F_{\alpha;\alpha_0}$.
In this sense, the state $\vert B(\alpha)\rangle_{\kappa}$ may be regarded as
the Ishibashi state of $F_{\alpha;\alpha_0}$.

We may now compare the expressions (\ref{eqn:ishiamp}) and (\ref{eqn:cbfsamp}) 
to find a possible free-field representation of the Ishibashi states of
minimal models.
Defining
\beq
\lishi (r,s)\vert = {}_{U}\langle a_{r,s}\vert+{}_{N}\langle a_{r,-s}\vert,
\label{eqn:ishibra}
\eeq
and 
\beq
\vert (r,s)\rishi = \vert a_{r,s}\rangle_{U}+\vert a_{r,-s}\rangle_{N},
\label{eqn:ishiket}
\eeq
with
\bea
&&{}_{U}\langle a_{r,s}\vert
=\sum_{k\in{\mathbb Z}}{}_{U}\langle B(\alpha_{r,s}+k\sqrt{pp'})\vert,
\label{eqn:mincoh1}\\
&&{}_{N}\langle a_{r,-s}\vert
=\sum_{k\in{\mathbb Z}}{}_{N}\langle B(\alpha_{r,-s}+k\sqrt{pp'})\vert,
\label{eqn:mincoh2}\\
&&\vert a_{r,s}\rangle_{U}
=\sum_{k\in{\mathbb Z}}\vert B(\alpha_{r,s}+k\sqrt{pp'})\rangle_{U},
\label{eqn:mincoh3}\\
&&\vert a_{r,-s}\rangle_{N}
=\sum_{k\in{\mathbb Z}}\vert B(\alpha_{r,-s}+k\sqrt{pp'})\rangle_{N},
\label{eqn:mincoh4}
\eea
where $\alpha_{r,s}$ are given by (\ref{eqn:minimalcharge}),
one can show that the states $\vert (r,s)\rishi$ diagonalise the overlaps and 
give minimal model characters,
\beq
\lishi (r,s)\vert(\tilde q^{1/2})^{L_0+\bar L_0-c/12}
\vert (r',s')\rishi=\delta_{rr'}\delta_{ss'}\chi_{(r,s)}(\tilde q).
\eeq
The appearance of the non-unitary sector even in unitary CFTs may seem odd, but this is necessary to describe cylinder diagrams where otherwise unphysical states would propagate. 
On the disk the non-unitary sector decouples and does not contribute to correlation functions
(Sec. III).
The states (\ref{eqn:ishibra}), (\ref{eqn:ishiket}) are a good 
candidate for the minimal model Ishibashi states as far as the modular 
properties are concerned.
We, however, immediately notice that such `Ishibashi' states are not unique
since $\vert (r,s)\rangle$ and $\vert (p'-r,p-s)\rangle$ give rise to the same
character but are perpendicular to each other.
In order to have a unique Ishibashi state for each primary field 
$(r,s)\sim(p'-r,p-s)$ of minimal models, we define the symmetrised states,
\bea
&&\lishi\phi_{r,s}\vert=\lishi\phi_{p'-r,p-s}\vert\nonumber\\
&&=\frac{1}{\sqrt 2}
(\lishi (r,s)\vert+\lishi (p'-r,p-s)\vert),
\label{eqn:symishibra}\\
%\eea
%\bea
&&\vert\phi_{r,s}\rishi=\vert\phi_{p'-r,p-s}\rishi\nonumber\\
&&=\frac{1}{\sqrt 2}
(\vert (r,s)\rishi+\vert (p'-r,p-s)\rishi),
\label{eqn:symishiket}
\eea
and shall regard them as our free-field realisation of the Ishibashi states.
Although this `symmetrisation' was not considered in \cite{coulomb},
such a prescription to ensure the equivalence of $(r,s)\sim(p'-r,p-s)$ is necessary.

\subsection{Cardy's consistent boundary states}

Physical boundary states in CFT are not only diffeomorphism invariant, but 
must also satisfy an extra constraint called Cardy's consistency condition.
We consider a cylinder of length $T$ and circumference $L$ as before 
and assume that boundary conditions $\tilde\alpha$ and 
$\tilde\beta$ are imposed on the two boundaries. 
Then, depending on how we define the direction of time, the partition function
on this cylinder may be written in two different ways. 
We may first regard the cylinder as an open string propagating in the periodic 
direction of time, with boundary conditions $\tilde\alpha$
and $\tilde\beta$ imposed at the two ends of the string. 
The partition function is then a sum of the chiral characters,
$Z_{\tilde\alpha\tilde\beta}(q)=\sum_jn^j_{\tilde\alpha\tilde\beta}\chi_j(q)$,
where $\chi_j(q)$ is the character for the chiral representation $j$ and 
$n^j_{\tilde\alpha\tilde\beta}$ is a non-negative integer representing the 
multiplicity of the representations.
We have defined $q=e^{-\pi L/T}$.
We may also see the cylinder as a closed string propagating from one boundary
(with boundary condition $\tilde\beta$) to the other (with $\tilde\alpha$).
Then the partition function is simply the cylinder amplitude between the two
boundaries, $\langle\tilde\alpha\vert e^{-TH}\vert\tilde\beta\rangle$.
Due to the equivalence of the two pictures, we have
$\sum_jn^j_{\tilde\alpha\tilde\beta}\chi_j(q)
=\langle\tilde\alpha\vert(\tilde q^{1/2})^{L_0+\tilde L_0-c/12}
\vert\tilde\beta\rangle$.
This is the consistency condition which needs to be satisfied by the boundary
states $\langle\tilde\alpha\vert$ and $\vert\tilde\beta\rangle$. 

If we have an appropriate basis of the boundary states, 
the right hand side of the consistency equation may be expanded using
the basis states $\{\langle a\vert\}$, $\{\vert b\rangle\}$ as
\beq
\sum_jn^j_{\tilde\alpha\tilde\beta}\chi_j(q)
=\sum_{a,b}\langle\tilde\alpha\vert a\rangle
\langle a\vert(\tilde q^{1/2})^{L_0+\bar L_0-c/12}\vert b\rangle
\langle b\vert\tilde\beta\rangle.
\label{eqn:consistency}
\eeq
Solving this equation, the {\em consistent} boundary states 
$\{\vert\tilde\alpha\rangle\}$ are found as linear combinations of the basis 
states.
It is convenient to use the Ishibashi states for such a basis.
For diagonal minimal models, using the property (\ref{eqn:ishiamp}) of the 
Ishibashi states and the modular transformation of the characters 
$\chi_i(q)=\sum_j S_{ij}\chi_j(\tilde q)$ under 
$\tau\rightarrow\tilde\tau=-1/\tau$, we have, by equating the 
coefficients of the characters, 
\beq
\sum_in^i_{\tilde\alpha\tilde\beta}S_{ij}=\langle\tilde\alpha\vert j\rishi
\lishi j\vert\tilde\beta\rangle.
\label{eqn:cardycond}
\eeq
Assuming the existence of a state $\vert\tilde 0\rangle$ such that 
$n^i_{\tilde 0\tilde\alpha}
=n^i_{\tilde\alpha\tilde 0}=\delta^i_{\tilde\alpha}$,
(\ref{eqn:cardycond}) was solved by Cardy\cite{cardymodular} as,
\beq
\vert\tilde\alpha\rangle
=\sum_j\vert j\rishi\lishi j\vert\tilde\alpha\rangle
=\sum_j\frac{S_{\alpha j}}{\sqrt{S_{0j}}}\vert j\rishi.
\label{eqn:consistentbs}
\eeq
Now, as the minimal model Ishibashi states have been written in the free-field 
representation (\ref{eqn:symishibra}) (\ref{eqn:symishiket}), Cardy's 
consistent boundary states (\ref{eqn:consistentbs}) can also be expressed using
our coherent boundary states by substituting (\ref{eqn:symishiket}) into 
(\ref{eqn:consistentbs}).

For the convenience of later discussions, let us spell out these in the
specific example of the Ising model.
The Ising model is the simplest non-trivial minimal model having the central
charge $c=1/2$ and is characterised by the two coprime integers $p=4$ and 
$p'=3$. 
There are three operators, the identity $I$, the energy $\epsilon$ and
the spin $\sigma$, having the conformal dimensions $0$, $1/2$, and $1/16$,
respectively, and are identified in the Kac table as 
$I=\phi_{1,1}=\phi_{2,3}$, $\epsilon=\phi_{2,1}=\phi_{1,3}$ and 
$\sigma=\phi_{1,2}=\phi_{2,2}$.
Using the modular transformation matrices for the characters, Cardy's boundary
states are written as\cite{cardymodular}
\bea
\vert\tilde I\rangle = \vert\tilde 0\rangle
&=&2^{-1/2}\vert I\rishi+2^{-1/2}\vert\epsilon\rishi+2^{-1/4}\vert\sigma\rishi,
\label{eqn:ishifixed1}\\
\vert\tilde\epsilon\rangle
&=&2^{-1/2}\vert I\rishi+2^{-1/2}\vert\epsilon\rishi-2^{-1/4}\vert\sigma\rishi,
\label{eqn:ishifixed2}\\
\vert\tilde\sigma\rangle
&=&\vert I\rishi-\vert\epsilon\rishi.
\label{eqn:ishifree}
\eea
It is argued that the first two states correspond to the fixed (up and down) 
boundary conditions since they differ only by the sign of $\vert\sigma\rishi$ 
which is associated with the spin operator.
The last state (\ref{eqn:ishifree}) is then identified as the free boundary 
state.
Using (\ref{eqn:ishiket}) and (\ref{eqn:symishiket}), these states are written
in our free-field representation 
as
%\footnote{These are different from (93)-(95) of \cite{coulomb} since we have
%symmetrised the two states for each representation as 
%(\ref{eqn:symishibra}), (\ref{eqn:symishiket}).}
\bea
\vert\tilde I\rangle
&=&2^{-1}(\vert a_{1,1}\rangle_U+\vert a_{1,-1}\rangle_N
+\vert a_{2,3}\rangle_U+\vert a_{2,-3}\rangle_N\nonumber\\
&&+\vert a_{2,1}\rangle_U+\vert a_{2,-1}\rangle_N
+\vert a_{1,3}\rangle_U+\vert a_{1,-3}\rangle_N)\nonumber\\
&&+2^{-3/4}(\vert a_{1,2}\rangle_U+\vert a_{1,-2}\rangle_N
+\vert a_{2,2}\rangle_U+\vert a_{2,-2}\rangle_N),\nonumber\\
\label{eqn:cfixed1}
\eea
\bea
\vert\tilde\epsilon\rangle
&=&2^{-1}(\vert a_{1,1}\rangle_U+\vert a_{1,-1}\rangle_N
+\vert a_{2,3}\rangle_U+\vert a_{2,-3}\rangle_N\nonumber\\
&&+\vert a_{2,1}\rangle_U+\vert a_{2,-1}\rangle_N
+\vert a_{1,3}\rangle_U+\vert a_{1,-3}\rangle_N)\nonumber\\
&&-2^{-3/4}(\vert a_{1,2}\rangle_U+\vert a_{1,-2}\rangle_N
+\vert a_{2,2}\rangle_U+\vert a_{2,-2}\rangle_N),\nonumber\\
\label{eqn:cfixed2}\\
%\eea
%\bea
\vert\tilde\sigma\rangle
&=&2^{-1/2}(\vert a_{1,1}\rangle_U+\vert a_{1,-1}\rangle_N
+\vert a_{2,3}\rangle_U+\vert a_{2,-3}\rangle_N\nonumber\\
&&-\vert a_{2,1}\rangle_U-\vert a_{2,-1}\rangle_N
-\vert a_{1,3}\rangle_U-\vert a_{1,-3}\rangle_N),\nonumber\\
\label{eqn:cfree}
\eea
where the states on the right hand sides are defined by (\ref{eqn:mincoh3}),
(\ref{eqn:mincoh4}). 
They are superpositions of countably many coherent states with different
boundary charges.

%%%%%%%%%%%%%%%%%%%%%%%%%%%%%%%%%%%%%%%%%%%%%%%%%%%%%%%%%%%%%%%%%%%

\section{Boundary correlation functions}

Now let us discuss how to compute boundary correlation functions
in our Coulomb-gas picture.
After giving the general formalism, we shall focus on the one point 
function of $\phi_{r,s}(z,\bar z)$ and the two point function 
of $\phi_{1,2}(z,\bar z)$, and derive their explicit expressions on the unit 
disk.
Once correlators on the disk are obtained, it is straightforward to map them on 
the half plane.
At the end of this section, we compare our free-field approach
and OPE computation of boundary correlation functions.

\subsection{Screened vertex operators and BRST states}

In the CBFS language, correlation functions on the full plane are described as 
follows\cite{felder}.
We define chiral screened vertex operators $V_{r,s}^{m,n}(z)$ as
\bea
V_{r,s}^{m,n}(z)&=&\oint\prod_{i=1}^{m}du_i\prod_{j=1}^n dv_j 
V_{r,s}(z)V_+(u_1)\cdots V_+(u_m)\nonumber\\
&&\times V_-(v_1)\cdots V_-(v_n),
\eea
where for conciseness we have denoted $V_{\alpha_{r,s}}(z)$ as $V_{r,s}(z)$
and $V_{\alpha_{\pm}}(z)$ as $V_{\pm}(z)$, and 
the integration contours are those of Felder's, all going through $z$ and encircling 
the origin (Fig.1 of \cite{felder}). 
Such an operator is a primary field of conformal dimension $h_{r,s}$.
We also denote the chiral CBFS $F_{\alpha_{r,s};\alpha_0}$ as 
$F_{r,s}$. 
The operator $V_{r,s}^{m,n}(z)$ defines a map from one Fock space
to another,
\beq
V_{r,s}^{m,n}(z):\;\; F_{r_0,s_0}\rightarrow F_{r_0+r-2m-1,s_0+s-2n-1}.
\label{eqn:cbfsmap}
\eeq
The $p$-point correlator,
\beq
\langle 0;\alpha_0\vert V_{r_1,s_1}^{m_1,n_1}(z_1)\cdots
V_{r_p,s_p}^{m_p,n_p}(z_p)\vert 0;\alpha_0\rangle,
\label{eqn:ppf1}
\eeq
is then seen as a sequence of mappings,
\beq
F_{1,1}\rightarrow F_{r_p-2m_p,s_p-2n_p}\rightarrow\cdots,
\eeq
and the final state
$V_{r_1,s_1}^{m_1,n_1}(z_1)\cdots V_{r_p,s_p}^{m_p,n_p}(z_p)
\vert 0;\alpha_0\rangle$ must belong to $F_{1,1}$ in order to have a 
non-trivial inner product with $\langle 0;\alpha_0\vert\in F_{1,1}^*$
(the dual module of $F_{1,1}$).
The same $p$-point correlator may be expressed differently, as
\beq
\langle\alpha_{p'-1,p-1};\alpha_0\vert V_{r_1,s_1}^{m'_1,n'_1}(z_1)\cdots
V_{r_p,s_p}^{m'_p,n'_p}(z_p)\vert 0;\alpha_0\rangle,
\label{eqn:ppf2}
\eeq
but this is in fact proportional to (\ref{eqn:ppf1}).
A key object in this formalism is the BRST operator,
\beq
Q_r=\frac{e^{2\pi i\alpha_+^2 r}-1}{r(e^{2\pi i\alpha_+^2}-1)}
\oint\prod_{i=1}^{r}du_i V_+(u_1)\cdots V_+(u_r),
\eeq
which maps $F_{r,s}$ to $F_{-r,s}$. 
The BRST operator is nilpotent, $Q_r Q_{p'-r}=0$, and physical states are 
realised as the cohomology space, ${\rm Ker}Q_r/{\rm Im}Q_{p'-r}$. 
An important point which is evident in this picture is that all intermediate
states appearing in the correlator are BRST states since the vacuum 
$\vert 0;\alpha_0\rangle$ is a BRST state and the screened vertex operators 
map BRST states to BRST states\cite{felder}.

We shall combine the above machinery and the boundary states of the previous section 
to compute correlation functions on the unit disk.
The in-state of the correlators is the non-chiral vacuum (in the unitary 
sector) at the origin, $\vert 0 \rangle = \vert 0,0;\alpha_0\rangle_U$, which 
is a BRST state.
For the out-state, we choose a boundary state ${}_U\langle B(\alpha)\vert$ of
(\ref{eqn:cohbsbra}) with a fixed boundary charge $\alpha$. 
We only consider the unitary sector since the non-unitary sector does not give 
non-trivial inner products with the in-state vacuum.
The correlators are then obtained by inserting non-chiral screened vertex
operators, 
\beq
{\cal V}_{(r_i,s_i), (\bar r_i,\bar s_i)}^{(m_i,n_i), (\bar m_i,\bar n_i)}
(z_i,\bar z_i)=
V_{r_i,s_i}^{m_i,n_i}(z_i)\bar V_{\bar r_i,\bar s_i}^{\bar m_i,\bar n_i}
(\bar z_i),
\eeq
between ${}_U\langle B(\alpha)\vert$ and $\vert 0,0;\alpha_0\rangle_U$.
As we focus on diagonal theories, the antiholomorphic vertex operators have 
the same conformal dimensions as the holomorphic counterparts, i.e., either
$(\bar r_i,\bar s_i)=(r_i,s_i)$ or $(\bar r_i,\bar s_i)=(p'-r_i,p-s_i)$.
Note that, in this construction, all the intermediate states are manifestly
BRST invariant because no spurious states arise.
Actual boundary $p$-point correlation functions for physical boundary
conditions are obtained by summing the fixed boundary-charge correlators,
\beq
{}_U\langle B(\alpha)\vert
\prod_{i=1}^p
V_{r_i,s_i}^{m_i,n_i}(z_i)\bar V_{\bar r_i,\bar s_i}^{\bar m_i,\bar n_i}
(\bar z_i)
\vert 0,0;\alpha_0\rangle_U,
\label{eqn:fbcc}
\eeq
over the boundary charges according to the linear combinations (such as
(\ref{eqn:cfixed1}) - (\ref{eqn:cfree})) given by Cardy's method.
Note that (\ref{eqn:fbcc}) is non-vanishing only for certain configurations of screening charges.
The net holomorphic and antiholomorphic charges are respectively,
\beq
-\alpha+\sum_i \alpha_{r_i,s_i} +\sum_i m_i\alpha_++\sum_in_i\alpha_-,
\eeq
and
\beq
\alpha-2\alpha_0+\sum_i \alpha_{\bar r_i,\bar s_i}
+\sum_i\bar m_i\alpha_++\sum_i\bar n_i\alpha_-,
\eeq
which must vanish independently.
These charge neutrality conditions associate the allowed values of $\alpha$ with the numbers of
holomorphic and antiholomorphic screening operators.

In the computation of the expression (\ref{eqn:fbcc}),      
$V_{r,s}(z)\bar V_{r,s}(\bar z)$,~ 
$V_{r,s}(z)\bar V_{p'-r,p-s}(\bar z)$,~ 
$V_{p'-r,p-s}(z)\bar V_{r,s}(\bar z)$ and 
$V_{p'-r,p-s}(z)\bar V_{p'-r,p-s}(\bar z)$ 
all correspond to a non-chiral field $\phi_{r,s}(z,\bar z)$ and one may use whichever combinations one wishes to use.
This is ensured by the fact that (\ref{eqn:fbcc}) is essentially a chiral $2p$-point function where the equivalence of $(r,s)\leftrightarrow(p'-r,p-s)$ (after the truncation of unphysical states) is guaranteed\cite{felder,dotfat}.
In particular, we are allowed to use $V_{r,s}(z)$ as the holomorphic and 
$\bar V_{p'-r, p-s}(\bar z)$ as the antiholomorphic (`mirror image') part of a single non-chiral operator (the analogy of a mirror and a boundary of CFT is based on the conformal Ward identity\cite{cardycorrel} which does not distinguish $\bar V_{r,s}(\bar z)$ from 
$\bar V_{p'-r, p-s}(\bar z)$).
Due to this equivalence, apparently different choices of vertex operators should all lead to a same
conformal block function.
In practice, likewise to the case of the Coulomb-gas computation without boundary, we shall choose such vertex operators that the number of screening operators is minimised. 
In the following subsections we shall evaluate the expression (\ref{eqn:fbcc}) for particular
cases of one and two point functions.

\subsection{Boundary one point functions}

For evaluation of the boundary one point correlator,
\beq
{}_U\langle B(\alpha)\vert
V_{r,s}^{m,n}(z)\bar V_{\bar r,\bar s}^{\bar m,\bar n}
(\bar z)
\vert 0,0;\alpha_0\rangle_U,
\label{eqn:b1pfunc}
\eeq
it is convenient to choose $(\bar r,\bar s)=(p'-r,p-s)$
(the other choice $(\bar r,\bar s)=(r,s)$ should lead to the same result but involves complicated
integral expressions).
According to (\ref{eqn:cbfsmap}), the holomorphic part of the CBFS is mapped as
\beq
F_{1,1}\rightarrow F_{r-2m,s-2n},
\eeq
and the antiholomorphic part is mapped as
\beq
\bar F_{1,1}\rightarrow \bar F_{p'-r-2\bar m,p-s-2\bar n}.
\eeq
Since ${}_U\langle B(\alpha)\vert\in F_{\alpha;\alpha_0}^*\otimes
F_{2\alpha_0-\alpha;\alpha_0}^*$, the correlator is non-vanishing only when
$F_{\alpha;\alpha_0}=F_{r-2m,s-2n}$ and 
$\bar F_{2\alpha_0-\alpha;\alpha_0}=\bar F_{p'-r-2\bar m,p-s-2\bar n}$,
that is,
\beq
\alpha=\frac 12(1-r+2m)\alpha_+
+\frac 12(1-s+2n)\alpha_-,
\label{eqn:b1pfcond1}
\eeq
and
\beq
2\alpha_0-\alpha=\frac 12(1-p'+r+2\bar m)\alpha_+
+\frac 12(1-p+s+2\bar n)\alpha_-.
\label{eqn:b1pfcond2}
\eeq
Summing (\ref{eqn:b1pfcond1}), (\ref{eqn:b1pfcond2})
and using $\alpha_+=\sqrt{p/p'}$, $\alpha_-=-\sqrt{p'/p}$, 
we have
$(m+\bar m)\alpha_++(n+\bar n)\alpha_-=0$,
or 
\beq
m+\bar m=0,\;\;\;n+\bar n=0,
\eeq
implying no screening charges. 
Then from (\ref{eqn:b1pfcond1}) we have $\alpha=\alpha_{r,s}$, and the
correlator (\ref{eqn:b1pfunc}) is evaluated as
\beq
{}_U\langle B(\alpha_{r,s})\vert
V_{r,s}^{0,0}(z)\bar V_{\bar r,\bar s}^{0,0}
(\bar z)\vert 0,0;\alpha_0\rangle_U
=(1-z\bar z)^{-2h},
\label{eqn:b1pfunc0}
\eeq
where $h=\alpha_{r,s}(\alpha_{r,s}-2\alpha_0)$.
As we shall see later, this is proportional to the two point correlator on the
full plane.

\begin{figure}
(a) Conformal block $I_{\rm I}$.\\\\
\epsfxsize=75mm
\epsfysize=60mm
\epsffile{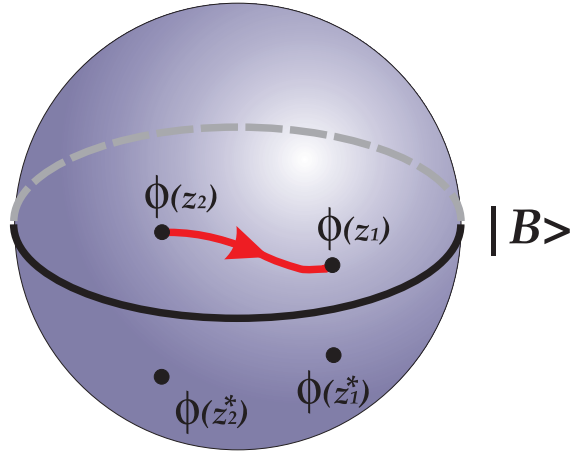}\\
\vspace{3mm}

(b) Conformal block $I_{\rm II}$.\\\\
\epsfxsize=75mm
\epsfysize=60mm
\epsffile{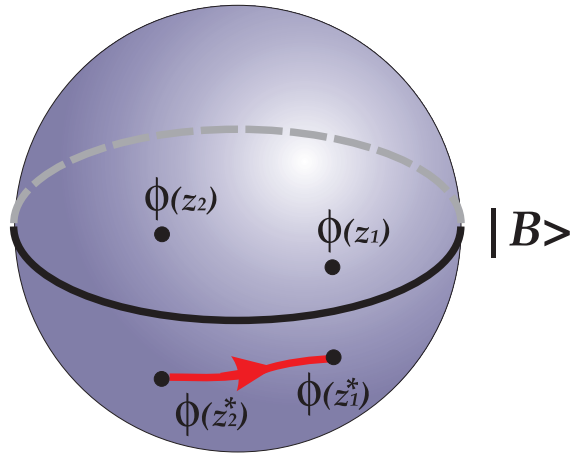}\\
\caption{\label{fig:epsart} 
The two conformal blocks $I_{\rm I}$ and $I_{\rm II}$ in the 
boundary two point function of $\phi_{1,2}$ on the disk.
The sphere represents the double of the disk, and the upper and lower 
hemispheres stand for the holomorphic and antiholomorphic sectors, which are 
glued on the boundary (the equator).
The lower hemisphere coordinates $z_i^*$ are obtained from $\bar z_i$ by the 
doubling $(z_i,\bar z_i)\rightarrow(z_i,z_i^*)$.
In the case of $I_{\rm I}$ where the screening operator $Q_-$ lie in 
the holomorphic sector, the integration contour can be deformed into the 
Pochhammer type around $z_1$ and $z_2$.
The integral is then proportional to the one from $z_2$ to $z_1$ (a).
Similarly, the screening contour of $I_{\rm II}$ is in the antiholomorphic 
sector and the integral is proportional to the one from $z_2^*$ to $z_1^*$ (b).
}
\end{figure}

\subsection{Boundary two point functions of $\phi_{1,2}$}

As a less trivial case, we consider the two point correlator of the primary
field $\phi_{1,2}$.
For the convenience of calculation we define one of the operators as
$\phi_{1,2}(z_1,\bar z_1)=V_{1,2}(z_1)\bar V_{p'-1,p-2}(\bar z_1)$ and the
other as $\phi_{1,2}(z_2,\bar z_2)=V_{1,2}(z_2)\bar V_{1,2}(\bar z_2)$.
The expression (\ref{eqn:fbcc}) then becomes
\bea
&&{}_U\langle B(\alpha)\vert
V_{1,2}^{m_1,n_1}(z_1)\bar V_{p'-1,p-2}^{\bar m_1,\bar n_1}(\bar z_1)
\nonumber\\
&&\times V_{1,2}^{m_2,n_2}(z_2)\bar V_{1,2}^{\bar m_2,\bar n_2}(\bar z_2)
\vert 0,0;\alpha_0\rangle_U.
\label{eqn:b2pfunc}
\eea
In the holomorphic and antiholomorphic sectors, the CBFSs are mapped as
\bea
&&F_{1,1}\rightarrow F_{1-2m_2,2-2n_2}\rightarrow F_{1-2m_1-2m_2,3-2n_1-2n_2},\\
&&\bar F_{1,1}\rightarrow\bar F_{1-2\bar m_2,2-2\bar n_2}\rightarrow
\bar F_{p'-2\bar m_1-2\bar m_2-1,p-2\bar n_1-2\bar n_2-1},\nonumber\\
&&
\eea
and hence, in order that the correlator be non-vanishing we must have
\bea
&&\alpha=(m_1+m_2)\alpha_++(n_1+n_2-1)\alpha_-,
\label{eqn:b2pfalpha}\\
&&2\alpha_0-\alpha=(\bar m_1+\bar m_2+1)\alpha_++(\bar n_1+\bar n_2+1)\alpha_-.
\nonumber\\
&&
\eea
Adding the above two expressions  we have
$(m+\bar m)\alpha_++(n+\bar n-1)\alpha_-=0$,
where $m=m_1+m_2$ and $n=n_1+n_2$ are the numbers of positive 
($\alpha_+$) and negative ($\alpha_-$) screening operators in the holomorphic 
sector, and similarly $\bar m=\bar m_1+\bar m_2$ and 
$\bar n=\bar n_1+\bar n_2$ for the antiholomorphic counterparts. 
This charge neutrality condition implies
\beq
m+\bar m=0,\;\;\; n+\bar n=1.
\label{eqn:b2pfcond}
\eeq
Then we have two possibilities:
\bea
({\rm I})&\;\;& m=\bar m=\bar n=0,\;\; n=1,\\
({\rm II})&\;\;& m=\bar m=n=0,\;\; \bar n=1.
\eea
From (\ref{eqn:b2pfalpha}), we have the boundary charge 
$\alpha=\alpha_{1,1}=0$ for (I) and 
$\alpha=\alpha_{1,3}=-\alpha_-$ for (II).

For the charge configuration (I), we have one screening operator 
\beq
Q_-=\oint dv V_-(v),
\eeq
in the holomorphic sector and thus the correlator (\ref{eqn:b2pfunc}) takes the
form,
\bea
I_{\rm I}&=&{}_U\langle B(\alpha_{1,1})\vert\oint dv
V_{1,2}(z_1)\bar V_{p'-1,p-2}(\bar z_1)V_-(v)\nonumber\\
&&{}\;\;
\times V_{1,2}(z_2)\bar V_{1,2}(\bar z_2)\vert 0,0;\alpha_0\rangle_U,
\label{eqn:b2pfunc1}
\eea
which, by an explicit calculation, reduces to
\bea
&&\oint dv (1-z_1\bar z_1)^a (1-z_1\bar z_2)^b
(1-v \bar z_1)^c (1-v \bar z_2)^d \nonumber\\
&&\;\;\; \times (1-z_2\bar z_1)^a (1-z_2\bar z_2)^b (z_1-v)^d 
\nonumber\\
&&\;\;\; \times (z_1-z_2)^b (v-z_2)^d (\bar z_1-\bar z_2)^a,
\eea
with $a=2\alpha_{1,2}(2\alpha_0-\alpha_{1,2})$, $b=2\alpha_{1,2}^2$,
$c=2\alpha_-(2\alpha_0-\alpha_{1,2})$, and
$d=2\alpha_-\alpha_{1,2}$.
As this expression is analytic, we may deform the integration contour as long 
as it is closed and non-contractable.
In this case the screening operator must lie entirely on the holomorphic part
and Felder's contour can be deformed into the Pochhammer type, going around 
$z_1$ and $z_2$.
The correlator is then proportional to the integration from $z_2$ to
$z_1$, and is written as
\bea
I_{\rm I}&=& {\cal N}_{\rm I}(1-z_1\bar z_1)^a (1-z_1\bar z_2)^b 
(1-z_2\bar z_1)^a\nonumber\\
&&\;\;\; \times  (1-z_2\bar z_2)^b (z_1-z_2)^b (\bar z_1-\bar z_2)^a\nonumber\\
&\times& \int_{z_2}^{z_1} dv
(1-v \bar z_1)^c (1-v \bar z_2)^d (z_1-v)^d (v-z_2)^d.\nonumber\\
&&
\eea
Here, ${\cal N}_{\rm I}$ is a constant arising from the deformation of the contour.
Note that, at this point, the expression is similar (in fact, proportional)
to the integral representation of a chiral four point conformal
block \cite{dotfat} (without boundary).
One may then proceed in the standard manner, namely, by fixing the projective
$SL(2,{\mathbb C})$ gauge, performing the integration, and then recovering the
coordinate dependence. We thus have,
\bea
I_{\rm I}&=&{\cal N}_{\rm I}(1-z_1\bar z_1)^a (1-z_2\bar z_2)^a
[\eta (\eta-1)]^a\nonumber\\
&\times&\frac{\Gamma(1-\alpha_-^2)^2}{\Gamma(2-2\alpha_-^2)}
F(2a,1-\alpha_-^2,2-2\alpha_-^2;\eta),
\label{eqn:I1}
\eea
where $F={}_2F_1$ is the hypergeometric function of the Gaussian type, and 
$\eta$ is defined as
\beq
\eta=\frac{(z_1-z_2)(\bar z_2-\bar z_1)}{(1-z_1\bar z_1)(1-z_2\bar z_2)}.
\label{eqn:eta}
\eeq

The calculation for (II) goes similarly. 
As we have one screening operator
\beq
\bar Q_-=\oint d\bar v \bar V_-(\bar v),
\eeq
in the antiholomorphic sector, the correlator is written as
\bea
I_{\rm II}&=&{}_U\langle B(\alpha_{1,3})\vert\oint d\bar v
V_{1,2}(z_1)\bar V_{p'-1,p-2}(\bar z_1)\bar V_-(\bar v)\nonumber\\
&&{}\;\;
\times V_{1,2}(z_2)\bar V_{1,2}(\bar z_2)\vert 0,0;\alpha_0\rangle_U\nonumber\\
&=& {\cal N}_{\rm II}(z_1-z_2)^b (\bar z_1-\bar z_2)^a (1-z_1\bar z_1)^a\nonumber\\
&&\;\;\; \times (1-z_1\bar z_2)^b (1-z_2\bar z_1)^a 
(1-z_2\bar z_2)^b  \nonumber\\
&\times& \int_{\bar z_2}^{\bar z_1} d\bar v
(\bar z_1-\bar v)^c (\bar v-\bar z_2)^d (1-z_1\bar v)^d (1-z_2\bar v)^d.
\nonumber\\
&&
\label{eqn:b2pfunc2}
\eea
We have again deformed Felder's integration contour (this time in the antiholomorphic sector)
into the Pochhammer contour around $\bar z_1$ and $\bar z_2$.
The resulting integral is proportional to the one from $\bar z_2$ to $\bar z_1$,
and ${\cal N}_{\rm II}$ is a constant.
Performing the integration we have
\bea
I_{\rm II}&=&{\cal N}_{\rm II}(1-z_1\bar z_1)^a (1-z_2\bar z_2)^a
[\eta (\eta-1)]^a(-\eta)^{b-a}\nonumber\\
&\times&\frac{\Gamma(1-\alpha_-^2)\Gamma(3\alpha_-^2-1)}{\Gamma(2\alpha_-^2)}
\nonumber\\
&\times&F(\alpha_-^2,1-\alpha_-^2,2\alpha_-^2;\eta).
\label{eqn:I2}
\eea
The two correlators $I_{\rm I}$ and $I_{\rm II}$ with fixed boundary 
charges are represented schematically (in the Schottky double picture) in 
Fig.1. 
They correspond to the two conformal blocks of the chiral four point function.

\subsection{Correlation functions on the half plane}

Boundary correlation functions obtained on the unit disk are mapped on the
half plane by the global conformal transformation,
\beq
w=-iy_0 \frac{z-1}{z+1},\;\;\;\bar w=iy_0\frac{\bar z-1}{\bar z+1},
\label{eqn:disktohp}
\eeq
which takes the unit circle $\vert z\vert=1$ on the $z$-plane to the infinite
line ${\rm Im} w=0$ on the $w$-plane, and the origin $z=0$ to the point
$w=iy_0$, $y_0\in {\mathbb R}$.
Under this transformation the holomorphic coordinate dependence on the 
unit disk is mapped on to the upper half $w$-plane, and the antiholomorphic 
dependence is mapped on to the lower half $\bar w$-plane. 
Using the transformation (\ref{eqn:disktohp}), $p$-point correlation functions 
on the half plane are written using those on the $z$-plane, 
i.e. on the disk, as
\bea
&&\langle\phi_1(w_1,\bar w_1)\cdots\phi_p(w_p,\bar w_p)\rangle_{\rm UHP}
\nonumber\\
&&=\prod_{i=1}^p
\left(\frac{dw_i}{dz_i}\right)^{-h_i}
\left(\frac{d\bar w_i}{d\bar z_i}\right)^{-\bar h_i}\!\!\!
\langle\phi_1(z_1,\bar z_1)\cdots\phi_p(z_p,\bar z_p)\rangle_{\rm disk}
\nonumber\\
&&=\prod_{i=1}^p
\left\{\frac{2y_0}{(z_i+1)(\bar z_i+1)}\right\}^{-2h_i}\!\!\!\!
\langle\phi_1(z_1,\bar z_1)\cdots\phi_p(z_p,\bar z_p)\rangle_{\rm disk},
\nonumber\\
&&
\eea
where we have assumed $h_i=\bar h_i$.
The parameter $\eta$ of (\ref{eqn:eta}) is mapped under this transformation
as
\beq
\eta=\frac{(z_1-z_2)(\bar z_2-\bar z_1)}{(1-z_1\bar z_1)(1-z_2\bar z_2)}
=\frac{(w_1-w_2)(\bar w_1-\bar w_2)}{(w_1-\bar w_1)(w_2-\bar w_2)},
\eeq
which is an anharmonic ratio of the four points $w_1$, $w_2$, $\bar w_1$ and
$\bar w_2$. 

Now, the boundary one point function of $\phi_{r,s}$ on the upper
half plane is easily found by using (\ref{eqn:b1pfunc0}) as
\bea
\langle\phi_{r,s}(w,\bar w)\rangle_{B(\alpha_{r,s})}
&=&\left\{\frac{2y_0(1-z\bar z)}{(z+1)(\bar z+1)}\right\}^{-2h}\nonumber\\
&=&[-i(w-\bar w)]^{-2h}\nonumber\\
&=&(2y)^{-2h},
\label{eqn:b1pfhp0}
\eea
where $h=h_{r,s}=\alpha_{r,s}(\alpha_{r,s}-2\alpha_0)$ is the conformal
dimension of the operator $\phi_{r,s}$, and $w=x+iy$, $\bar w=w^*=x-iy$,
$x,y\in {\mathbb R}$.
Note that the result is $y_0$-independent.
The two point function of $\phi_{1,2}$ on the disk is mapped on to the half 
plane similarly.
For the conformal block (I) we have
\bea
&&\langle\phi_{1,2}(w_1,\bar w_1)\phi_{1,2}(w_2,\bar w_2)
\rangle_{B(\alpha_{1,1})}\nonumber\\
&=&\left\{\frac{4y_0^2}{(z_1+1)(\bar z_1+1)(z_2+1)(\bar z_2+1)}\right\}^{-2h}
I_{\rm I}\nonumber\\
&=&{\cal N}_{\rm I}
\left\{\frac{(w_1-\bar w_1)(\bar w_2-w_2)}{(w_1-w_2)(\bar w_1-\bar w_2)
(w_1-\bar w_2)(\bar w_1-w_2)}\right\}^{2h}\nonumber\\
&&\times \frac{\Gamma(1-\alpha_-^2)^2}{\Gamma(2-2\alpha_-^2)}
F(-4h,1-\alpha_-^2,2-2\alpha_-^2;\eta),
\label{eqn:b2pfhp1}
\eea
and for (II) we have
\bea
&&\langle\phi_{1,2}(w_1,\bar w_1)\phi_{1,2}(w_2,\bar w_2)
\rangle_{B(\alpha_{1,3})}\nonumber\\
&=&\left\{\frac{4y_0^2}{(z_1+1)(\bar z_1+1)(z_2+1)(\bar z_2+1)}\right\}^{-2h}
I_{\rm II}\nonumber\\
&=&{\cal N}_{\rm II}
\left\{\frac{(w_1-\bar w_1)(\bar w_2-w_2)}{(w_1-w_2)(\bar w_1-\bar w_2)
(w_1-\bar w_2)(\bar w_1-w_2)}\right\}^{2h}\nonumber\\
&&\times \frac{\Gamma(1-\alpha_-^2)\Gamma(3\alpha_-^2-1)}{\Gamma(2\alpha_-^2)}
(-\eta)^{2h+2\alpha_-^2}\nonumber\\
&&\times F(\alpha_-^2,1-\alpha_-^2,2\alpha_-^2;\eta),
\label{eqn:b2pfhp2}
\eea
where $h=h_{1,2}=\alpha_{1,2}(\alpha_{1,2}-2\alpha_0)$. 
Physical correlation functions are linear sums of these conformal blocks 
where the coefficients are given by Cardy's states.

\subsection{Boundary states and conformal blocks}
Before illustrating in specific examples, we mention how the above description of boundary correlation functions fits into the conventional discussion of \cite{cardylewellen,lewellen}, and
see the validity and limitation of the Coulomb-gas approach.

The charge neutrality conditions (\ref{eqn:b1pfcond1}) (\ref{eqn:b1pfcond2}) for a one-point function pick up 
the coefficient of the corresponding Ishibashi state from a Cardy state, since $\alpha=\alpha_{r,s}$ is the only
boundary charge which gives a non-vanishing term.
This agrees with our understanding that the coefficients of Cardy's state are essentially the one-point coupling constants of bulk (closed string vertex) operators to the boundary (brane)\cite{cardylewellen,lewellen}.
Once one-point coupling constants are known, it is in principle possible to compute boundary
multipoint functions since they reduce to one point functions after repeated use of bulk OPEs (Fig.2a).
In particular, we may start such a procedure from the farthest point from the boundary (in the 
radial ordering sense), approaching the boundary by performing OPE with the farthest remaining point one by one.
Due to naturality of CFT, such OPEs are translated into the fusions of operators, 
\bea
&&[\phi_1]\times [\phi_2]=[j_1],\nonumber\\
&&[j_1]  \times [\phi_3]=[j_2],\nonumber\\
&&\cdots\nonumber\\
&&[j_{p-2}]\times [\phi_p]=[j_{p-1}],
\eea
which define a conformal block with no subchains (Fig.2b).
As the fusion of primary operators is equivalent to the map (\ref{eqn:cbfsmap}) between CBFSs restricted to 
BRST subspaces\cite{felder}, the conformal block of Fig.2b is represented by our fixed 
boundary-charge correlator (\ref{eqn:fbcc}) with $\alpha$ corresponding to $[j_{p-1}]$.
The Ishibashi state $\lishi j_{p-1}\vert$ acts as a filter (or a half mirror) transmitting only the 
Virasoro representation $[j_{p-1}]$. 
This property of Ishibashi states is captured by boundary charges and charge neutrality.

Due to the absence of internal channels, two point functions on the disk (or half plane) are completely determined by a boundary state, apart from the normalisation of conformal blocks.
In the case of $p$-point functions with $p\geq 3$, however, states in the internal channels 
$j_{i<p-1}$ cannot in general be determined uniquely even if the state in the `boundary channel' $j_{p-1}$ is fixed (an example is the spin three point function of the Ising model with $j_2=\sigma$, where $j_1$ can be $I$ or $\epsilon$).
Corresponding to this, the contours of (\ref{eqn:fbcc}) with $p\geq 3$ may be deformed in several different ways to give independent convergent functions which are expected to reproduce conformal blocks with different internal states.
The relative coefficients of such conformal blocks in a boundary correlation function cannot be determined by the boundary state (as these conformal blocks belong to the same Ishibashi state) but should be constrained by information of the bulk.
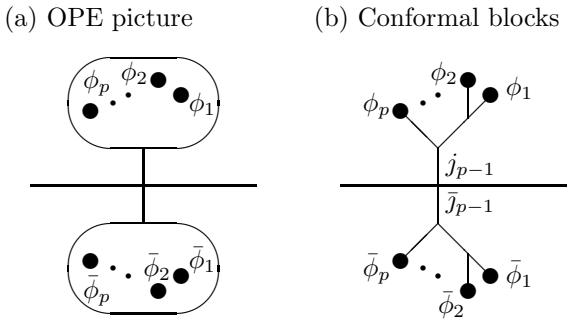
\begin{figure}

\setlength{\unitlength}{1mm}
\begin{center}
\begin{minipage}{40mm}
(a) OPE picture

~~~\begin{picture}(30,40)
{\thicklines
\put(0,20){\line(1,0){30}}}
{\thinlines
\put(15,25){\line(0,-1){10}}}
\put(8,30){\circle*{2}}
\put(20,32){\circle*{2}}
\put(17,34){\circle*{2}}
\put(8,10){\circle*{2}}
\put(20,8){\circle*{2}}
\put(17,6){\circle*{2}}
\put(11,31){\circle*{0.8}}
\put(13,32){\circle*{0.8}}
\put(11,9){\circle*{0.8}}
\put(13,8){\circle*{0.8}}
\put(7,33){$\phi_p$}
\put(21,30){$\phi_1$}
\put(12,34){$\phi_2$}
\put(7,5){$\bar\phi_p$}
\put(21,9){$\bar\phi_1$}
\put(15,8){$\bar\phi_2$}
\put(15,31){\oval(20,12)}
\put(15,9){\oval(20,12)}
\end{picture}
\end{minipage}
\begin{minipage}{40mm}
(b) Conformal blocks

~~~\begin{picture}(30,40)
{\thicklines
\put(0,20){\line(1,0){30}}}
{\thinlines
\put(13,25){\line(0,-1){10}}
\put(13,25){\line(1,1){7}}
\put(13,25){\line(-1,1){5}}
\put(17,29){\line(0,1){5}}
\put(13,15){\line(1,-1){7}}
\put(13,15){\line(-1,-1){5}}
\put(17,11){\line(0,-1){5}}}
\put(8,30){\circle*{2}}
\put(20,32){\circle*{2}}
\put(17,34){\circle*{2}}
\put(8,10){\circle*{2}}
\put(20,8){\circle*{2}}
\put(17,6){\circle*{2}}
\put(11,31){\circle*{0.8}}
\put(13,32){\circle*{0.8}}
\put(11,9){\circle*{0.8}}
\put(13,8){\circle*{0.8}}
\put(3,30){$\phi_p$}
\put(22,32){$\phi_1$}
\put(12,34){$\phi_2$}
\put(3,8){$\bar\phi_p$}
\put(22,7){$\bar\phi_1$}
\put(13,3){$\bar\phi_2$}
\put(14,22){$j_{p-1}$}
\put(14,17){$\bar\jmath_{p-1}$}
\end{picture}
\end{minipage}
\end{center}

\caption{OPE and conformal blocks of boundary $p$-point functions.
Repeating OPEs in the bulk, the boundary correlation function reduces to one point functions (a). 
This can also be seen as conformal blocks with internal channels $j_{i<p-1}$ and boundary channel
$j_{p-1}$ (b).}
\end{figure}

%%%%%%%%%%%%%%%%%%%%%%%%%%%%%%%%%%%%%%%%%%%%%%%%%%%%%%%%%%%%%%%%%%%

\section{Ising model}

We shall illustrate the method presented in the previous section in the
example of the critical Ising model. 
Before starting actual calculations we notice from the discussions of the
previous section that the non-unitary sector and the boundary states with 
charges outside the Kac table do not contribute to correlation functions. 
Neglecting such unnecessary terms in (\ref{eqn:cfixed1}) - (\ref{eqn:cfree}), 
for the bra-boundary states of the Ising model we have,
\bea
\langle\tilde I\vert
&\sim&2^{-1}(
{}_U\langle B(\alpha_{1,1})\vert+{}_U\langle B(\alpha_{2,3})\vert
\nonumber\\
&&+
{}_U\langle B(\alpha_{2,1})\vert+{}_U\langle B(\alpha_{1,3})\vert)\nonumber\\
&&+2^{-3/4}(
{}_U\langle B(\alpha_{1,2})\vert+{}_U\langle B(\alpha_{2,2})\vert),
\label{eqn:sfixed1}
\eea
\bea
\langle\tilde\epsilon\vert
&\sim&2^{-1}(
{}_U\langle B(\alpha_{1,1})\vert+{}_U\langle B(\alpha_{2,3})\vert
\nonumber\\
&&+
{}_U\langle B(\alpha_{2,1})\vert+{}_U\langle B(\alpha_{1,3})\vert)\nonumber\\
&&-2^{-3/4}({}_U\langle B(\alpha_{1,2})\vert+{}_U\langle B(\alpha_{2,2})\vert),
\label{eqn:sfixed2}
\eea
\bea
\langle\tilde\sigma\vert
&\sim&2^{-1/2}(
{}_U\langle B(\alpha_{1,1})\vert+{}_U\langle B(\alpha_{2,3})\vert
\nonumber\\
&&-{}_U\langle B(\alpha_{2,1})\vert-{}_U\langle B(\alpha_{1,3})\vert).
\label{eqn:sfree}
\eea
Since these are linear combinations of fixed boundary-charge states
${}_U\langle B(\alpha_{r,s})\vert$, the correlation functions on the disk with 
physical boundary conditions $\tilde I$, $\tilde\epsilon$ and $\tilde\sigma$ 
are given by linear combinations of fixed boundary-charge correlators 
(\ref{eqn:fbcc}).
Using the global conformal transformation explained in the last section,
we shall obtain correlation functions on the half plane and compare them with existing results.

\subsection{One point functions of spin and energy operators}

Let us first consider the spin one point function. 
We may choose either $\sigma=\phi_{1,2}$ or $\sigma=\phi_{2,2}$. 
Let, for definiteness, $(r,s)=(1,2)$ in the holomorphic part and 
in order to use the result of the last section, 
$(\bar r,\bar s)=(2,2)$ in the antiholomorphic part. 
From (\ref{eqn:b1pfunc0}) we immediately notice that only the state
${}_U\langle B(\alpha_{1,2})\vert$ contributes to the one point function, 
all other states giving vanishing correlators. 
For the boundary condition $\tilde I$, the one point function on the disk is
\bea
&&\langle\tilde I\vert \sigma(z,\bar z)\vert 0\rangle\nonumber\\
&&=\langle\tilde I\vert V_{1,2}(z)\bar V_{2,2}(\bar z)
\vert 0,0;\alpha_0\rangle_U\nonumber\\
&&=2^{-3/4}{}_U\langle B(\alpha_{1,2})
\vert V_{1,2}(z)\bar V_{2,2}(\bar z)\vert 0,0;\alpha_0\rangle_U\nonumber\\
&&=2^{-3/4}(1-z\bar z)^{-1/8}.
\eea
Properly normalised one point function is then,
\beq
\frac{\langle\tilde I\vert\sigma(z,\bar z)\vert 0\rangle}{\langle\tilde I\vert 0\rangle}
=2^{1/4}(1-z\bar z)^{-1/8}.
\eeq
This is mapped on to the half plane by using the conformal transformation
(\ref{eqn:b1pfhp0}), as
\beq
\langle\sigma(w,\bar w)\rangle_{\tilde I}
=\langle\sigma(y)\rangle_{\tilde I}
=2^{1/4}(2y)^{-1/8},
\eeq
where $y$ is the distance from the boundary. 
Likewise, boundary spin correlation functions for the conditions 
$\tilde\epsilon$ and $\tilde\sigma$ are obtained simply by picking up the 
coefficients of ${}_U\langle B(\alpha_{1,2})\vert$ in (\ref{eqn:sfixed2}) and 
(\ref{eqn:sfree}), and are normalised using $\langle\tilde\epsilon\vert 0\rangle=1/2$
and $\langle\tilde\sigma\vert 0\rangle=1/\sqrt 2$. 
On the half plane, they are
\beq
\langle\sigma(w,\bar w)\rangle_{\tilde\epsilon}
=\langle\sigma(y)\rangle_{\tilde\epsilon}
=-2^{1/4}(2y)^{-1/8},
\eeq
and
\beq
\langle\sigma(w,\bar w)\rangle_{\tilde\sigma}
=\langle\sigma(y)\rangle_{\tilde\sigma}
=0.
\eeq
Hence, $\tilde I$ and $\tilde\epsilon$ are indeed the fixed (up and down) and 
$\tilde\sigma$ is the free boundary condition, as is stated in 
\cite{cardymodular,cardylewellen}.
In our Coulomb-gas formalism the relation between one point functions and 
the coefficients of Ishibashi states is explained by the neutrality of charges.

Putting, say, $(r,s)=(2,1)$ and $(\bar r,\bar s)=(1,3)$, the energy one point
function is obtained similarly.
On the half plane we have,
\bea
&&\langle\epsilon(y)\rangle_{\tilde I}
=\langle\epsilon(y)\rangle_{\tilde\epsilon}=(2y)^{-1},\\
&&\langle\epsilon(y)\rangle_{\tilde\sigma}=-(2y)^{-1}.
\eea

\subsection{Spin two point function}

Next, let us consider the spin two point function. 
Since $\sigma=\phi_{1,2}$, we can use the result of Subsec.C of the preceding 
section.
There are only two values of boundary charges, $-\alpha_{1,1}$ and 
$-\alpha_{1,3}$, which give non-trivial contributions to the correlator. 
The two corresponding states ${}_U\langle B(\alpha_{1,1})\vert$ and 
${}_U\langle B(\alpha_{1,3})\vert$ give rise to the two conformal blocks
$I_{\rm I}$ and $I_{\rm II}$, respectively, and the correlation function is a
linear combination of these conformal blocks with coefficients
given by Cardy's states (\ref{eqn:sfixed1}) - (\ref{eqn:sfree}).
Then,
\bea
\frac{\langle\tilde I\vert \sigma(z_1,\bar z_1)\sigma(z_2,\bar z_2)\vert 0\rangle}
{\langle\tilde I\vert 0\rangle}
&=&\frac{\langle\tilde\epsilon\vert \sigma(z_1,\bar z_1)\sigma(z_2,\bar z_2)\vert 0\rangle}
{\langle\tilde\epsilon\vert 0\rangle}
\nonumber\\
&=&I_{\rm I}+I_{\rm II},\\
\frac{\langle\tilde\sigma\vert \sigma(z_1,\bar z_1)\sigma(z_2,\bar z_2)\vert 0\rangle}
{\langle\tilde\sigma\vert 0\rangle}&=&I_{\rm I}-I_{\rm II},
\eea
where the actual forms of $I_{\rm I}$ and $I_{\rm II}$ are given by
(\ref{eqn:I1}) and (\ref{eqn:I2}), with $a=-1/8$,
$b=3/8$ and $\alpha_-^2=3/4$. 
In this case the hypergeometric functions reduce to,
\bea
&&{}_2F_1(-\frac 14,\frac 14,\frac 12;\eta)
=\frac{\sqrt{1+\sqrt{1-\eta}}}{\sqrt 2},
\label{eqn:hypergeom1}\\
&&{}_2F_1(\frac 34,\frac 14,\frac 32;\eta)
=\frac{\sqrt{2(1-\sqrt{1-\eta})}}{\sqrt \eta}.
\label{eqn:hypergeom2}
\eea
Using (\ref{eqn:hypergeom1}), (\ref{eqn:hypergeom2}) and the conformal
transformation (\ref{eqn:disktohp}), we find the two point functions on the
half plane,
\bea
&&\langle\sigma(w_1,\bar w_1)\sigma(w_2,\bar w_2)\rangle_{\tilde I}\nonumber\\
&=&\langle\sigma(w_1,\bar w_1)\sigma(w_2,\bar w_2)\rangle_{\tilde\epsilon}
\nonumber\\
&=&\frac{\tilde{\cal N}}{\sqrt 2}
\left\{\frac{(w_1-\bar w_1)(\bar w_2-w_2)}{(w_1-w_2)(\bar w_1-\bar w_2)
(w_1-\bar w_2)(\bar w_1-w_2)}\right\}^{1/8}\nonumber\\
&&\times\left({\cal N}_{\rm I}\sqrt{\sqrt{1-\eta}+1}+{\cal N}_{\rm II}\sqrt{\sqrt{1-\eta}-1}\right),\\
%\eea
%\bea
&&\langle\sigma(w_1,\bar w_1)\sigma(w_2,\bar w_2)\rangle_{\tilde\sigma}
\nonumber\\
&=&\frac{\tilde{\cal N}}{\sqrt 2}
\left\{\frac{(w_1-\bar w_1)(\bar w_2-w_2)}{(w_1-w_2)(\bar w_1-\bar w_2)
(w_1-\bar w_2)(\bar w_1-w_2)}\right\}^{1/8}\nonumber\\
&&\times\left({\cal N}_{\rm I}\sqrt{\sqrt{1-\eta}+1}-{\cal N}_{\rm II}\sqrt{\sqrt{1-\eta}-1}\right),
\eea
where ${\tilde{\cal N}}=\Gamma(1/4)^2/\Gamma(1/2)$. 
Studying the behaviours away from the boundary (we accept the convention that two point functions of bulk operators are normalised as 
$\langle\phi_i(w_1)\phi_j(w_2)\rangle=\delta_{ij}(w_1-w_2)^{-2h_i}$) and comparing the leading terms with the OPE coefficients of \cite{lewellen},
the normalisation of the conformal blocks are fixed as
${\cal N}_{\rm I}={\cal N}_{\rm II}=\Gamma(1/2)/\Gamma(1/4)^2$.

This result was obtained long time ago\cite{cardycorrel}, by solving a 
differential equation to find the two conformal blocks $I_{\rm I}$, 
$I_{\rm II}$, and then fixing the coefficients by considering 
asymptotic behaviours of the correlation function.
The (relative) coefficients of the conformal blocks are now attributed to the coefficients of Cardy's states in our Coulomb-gas approach, although we have used the asymptotic behaviours to fix
the normalisation of each conformal block.

\subsection{Energy two point function}

Finally we derive the energy two point function in our formalism.
As $\epsilon=\phi_{2,1}$, the calculation is parallel to the case of the
spin two point function. 
From the charge neutrality condition we find that non-vanishing correlators
arise from the states ${}_U\langle B(\alpha_{1,1})\vert$ and 
${}_U\langle B(\alpha_{3,1})\vert$.
However, as none of the boundary states (\ref{eqn:sfixed1}) - (\ref{eqn:sfree})
contain ${}_U\langle B(\alpha_{3,1})\vert$, only
${}_U\langle B(\alpha_{1,1})\vert$ gives non-trivial contribution to the
correlation function.
Hence, the energy two point function does not depend on boundary conditions.
After a simple calculation we find, on the half plane,
\bea
&&\langle\epsilon(w_1,\bar w_1)\epsilon(w_2,\bar w_2)\rangle_{\tilde I, \tilde\epsilon, \tilde\sigma}
\nonumber\\
&=&{\cal N}
\frac{(w_1-\bar w_1)(\bar w_2-w_2)}{(w_1-w_2)(\bar w_1-\bar w_2)
(w_1-\bar w_2)(\bar w_1-w_2)}\nonumber\\
&\times&\frac{\Gamma(-1/3)^2}{\Gamma(-2/3)}F(-2,-1/3,-2/3;\eta).
\label{eqn:energy2pf}
\eea
The normalisation constant ${\cal N}$ is determined as ${\cal N}=\Gamma(-2/3)/\Gamma(-1/3)^2$,
by considering the off-boundary behaviour.
The hypergeometric function turns out to be an algebraic function
$F(-2,-1/3,-2/3;\eta)=1-\eta+\eta^2$, and using the coordinates
$w_i=x_i+iy_i$, $\bar w_i=w_i^*=x_i-iy_i$ ($x_i$, $y_i\in{\mathbb R}$),
the correlation function (\ref{eqn:energy2pf}) is written as
\bea
&&\langle\epsilon(x_1,y_1)\epsilon(x_2,y_2)\rangle_{\tilde I, \tilde\epsilon, \tilde\sigma}\nonumber\\
&&=\frac{4y_1y_2}{[(x_1-x_2)^2+(y_1-y_2)^2][(x_1-x_2)^2+(y_1+y_2)^2]}\nonumber\\
&&+\frac{1}{4y_1y_2}.
\eea
This agrees with the result in \cite{cardycorrel}.

%%%%%%%%%%%%%%%%%%%%%%%%%%%%%%%%%%%%%%%%%%%%%%%%%%%%%%%%%%%%%%%%%%%

\section{Summary}

In this paper we have described a novel method to calculate correlation 
functions of two dimensional CFT on the disk and on the half plane. 
We have used the free-field construction of boundary states developed in
\cite{coulomb}, and derived boundary correlation functions for the boundary 
states classified by Cardy's method.
The key feature of our formalism is the neutrality of bulk and boundary
charges, which associates the coefficients in Cardy's boundary states
directly with the linear combinations of conformal blocks.
Thus we could unify the two important pieces of boundary CFT, namely, boundary
correlation functions\cite{cardycorrel} and consistent boundary states of 
Cardy\cite{cardymodular}, by using the Coulomb-gas picture.
We have checked the formalism in the Ising model, and shown that our method
reproduces the known results.

Cardy's classification of boundary states has been generalised by Lewellen
\cite{lewellen} and Pradisi, Sagnotti, Stanev\cite{pradisi} beyond the 
diagonal models, and the Coulomb-gas technique is also known to be applicable 
to more general CFTs, such as WZNW models\cite{gmoms} and CFTs with W-algebra
\cite{bouwknegt,fl,mizoguchi}.
We therefore expect that the method discussed in this paper may be
applied to such CFTs relatively easily.
In particular, from a string theory point of view, application to WZNW theories
seems to be quite fruitful since it would provide an alternative method to find
correlation functions with D-branes on a group manifold.
We shall discuss such issues in separate publications\cite{ckw}.

%%%%%%%%%%%%%%%%%%%%%%%%%%%%%%%%%%%%%%%%%%%%%%%%%%%%%%%%%%%%%%%%%%%

\begin{acknowledgments}

I thank Dr. John Wheater for useful discussions and constant encouragement.
I am also grateful to Prof. John Cardy, Dr. Matthias Gaberdiel
and Dr. Ian Kogan for helpful discussions.
\end{acknowledgments}

%%%%%%%%%%%%%%%%%%%%%%%%%%%%%%%%%%%%%%%%%%%%%%%%%%%%%%%%%%%%%%%%%%%

%\begin{thebibliography}{99}

%%%%%%%%%%%%%%%%%%%%%%%%%%%%%%%%%%%%%%%%%%%%%%%%%%%%%%%%%%%%%%%%%%%

\end{document}